\documentclass[preprint,showpacs,aps,amsmath,amssymb,
nofootinbib,superscriptaddress,showkeys]{revtex4}

\usepackage{graphicx,floatflt,epsfig,epsf} 

\def\bar {\overline}

\def\delgs {\Delta\Gamma_s}
\def\delms {\Delta M_s}
\def\delgd {\Delta\Gamma_d}
\def\delmd {\Delta M_d}

\def\be {\begin{equation}}
\def\ee {\end{equation}}
\def\bea {\begin{eqnarray}}
\def\eea {\end{eqnarray}}

\def\bra {\langle}
\def\ket {\rangle}

\def\bsbsbar{B_s$--$\bar{B}_s}
\def\bdbdbar{B_d$--$\bar{B}_d}
\def\bqbqbar{B_q$--$\bar{B}_q}
\def\dgd{\Delta \Gamma_d}
\def\dgs{\Delta \Gamma_s}
\def\dgq{\Delta \Gamma_q}
\def\gd{\Gamma_d}
\def\gs{\Gamma_s}
\def\gq{\Gamma_q}
\def\beq{\begin{equation}}
\def\eeq{\end{equation}}
\def\barr{\begin{eqnarray}}
\def\earr{\end{eqnarray}}

\def\opcit(#1){ {\em op. cit.}, #1}
\def\etal {\em et al.}

\def\issue(#1,#2,#3){#1 (#3) #2} 

\def\APP(#1,#2,#3){Acta Phys.\ Polon.\ \issue(#1,#2,#3)}
\def\ARNPS(#1,#2,#3){Ann.\ Rev.\ Nucl.\ Part.\ Sci.\ \issue(#1,#2,#3)}
\def\CPC(#1,#2,#3){Comp.\ Phys.\ Comm.\ \issue(#1,#2,#3)}
\def\CIP(#1,#2,#3){Comput.\ Phys.\ \issue(#1,#2,#3)}
\def\EPJC(#1,#2,#3){Eur.\ Phys.\ J.\ C\ \issue(#1,#2,#3)}
\def\EPJD(#1,#2,#3){Eur.\ Phys.\ J. Direct\ C\ \issue(#1,#2,#3)}
\def\IEEETNS(#1,#2,#3){IEEE Trans.\ Nucl.\ Sci.\ \issue(#1,#2,#3)}
\def\IJMP(#1,#2,#3){Int.\ J.\ Mod.\ Phys. \issue(#1,#2,#3)}
\def\JHEP(#1,#2,#3){J.\ High Energy Physics \issue(#1,#2,#3)}
\def\JPG(#1,#2,#3){J.\ Phys.\ G \issue(#1,#2,#3)}
\def\MPL(#1,#2,#3){Mod.\ Phys.\ Lett.\ \issue(#1,#2,#3)}
\def\NP(#1,#2,#3){Nucl.\ Phys.\ \issue(#1,#2,#3)}
\def\NIM(#1,#2,#3){Nucl.\ Instrum.\ Meth.\ \issue(#1,#2,#3)}
\def\PL(#1,#2,#3){Phys.\ Lett.\ \issue(#1,#2,#3)}
\def\PRD(#1,#2,#3){Phys.\ Rev.\ D \issue(#1,#2,#3)}
\def\PRL(#1,#2,#3){Phys.\ Rev.\ Lett.\ \issue(#1,#2,#3)}
\def\SJNP(#1,#2,#3){Sov.\ J. Nucl.\ Phys.\ \issue(#1,#2,#3)}
\def\ZPC(#1,#2,#3){Zeit.\ Phys.\ C \issue(#1,#2,#3)}

\begin{document}
  
\begin{flushright} 
CU-PHYSICS/08-2007\\
TIFR/TH/07-09\\
\end{flushright}

\title{Possibility of large lifetime differences in 
neutral B meson systems}

\author{Amol Dighe}
\affiliation{Tata Institute of Fundamental Research,\\
Homi Bhabha Road, Colaba, Mumbai 400005, India}
\author{Anirban Kundu}
\affiliation{Department of Physics, University of Calcutta,\\     
92 Acharya Prafulla Chandra Road, Kolkata 700009, India}
\author{Soumitra Nandi}
\affiliation{Department of Physics, University of Calcutta,\\     
92 Acharya Prafulla Chandra Road, Kolkata 700009, India}

\begin{abstract} 
We investigate new physics models that can increase the
lifetime differences in the $B_q$--$\bar{B}_q$ 
systems ($q = d,s$)
above their standard model values.
If both $B_q$ as well as $\bar{B}_q$ can decay to a final state through
flavour dependent new physics interactions, the 
so-called Grossman bound may be evaded.
As examples, we consider the scalar leptoquark
model and $\lambda''$-type R-parity violating supersymmetry.
We find that models with a scalar leptoquark can
enhance $\Delta\Gamma_s/\Gamma_s$ all the way up to
its experimental upper bound
and $\Delta\Gamma_d/\Gamma_d$ to as much as $\sim 2.5\%$,
at the same time allowing the CP violating phase
$\beta_s$ to vary between $- 45^\circ$ and $20^\circ$.
R-parity violating supersymmetry models 
cannot enhance the lifetime differences significantly,
but can enhance the value of $\beta_s$ up to
$\sim \pm 20^\circ$.
This may bring the values of $\Delta\Gamma_q/\Gamma_q$ 
as well as $\beta_s$ within the measurement capabilities 
of $B$ factories and LHCb.
We also obtain bounds on combinations of these new
physics couplings, and predict enhanced branching ratios 
of $B_{s/d} \to \tau^+ \tau^-$.
\end{abstract}



\maketitle

\section{Introduction}
\label{intro}

The standard model (SM) has been successful in explaining almost 
all the observations at the accelerator experiments so far, which
makes it a challenge to look for signals of new physics.
Apart from the direct searches for new particles at colliders,
tests of the low energy predictions of the SM can also provide
indirect signatures of the physics beyond the standard model
(BSM).
In the domain of flavour physics in particular, such low
energy observables include the branching ratios of various
$B$ decay modes, the extent of CP violation in these decays,
as well as the oscillation parameters of neutral $B$ meson 
systems \cite{bf-review}. 
The data from the $B$ factories and the Tevatron
have already played a crucial role in constraining the
nature and extent of BSM physics.

In this paper we shall concentrate on the oscillation
parameters in the $\bdbdbar$ as well as $\bsbsbar$ systems.
For convenience, we shall refer to the labels 
$d$ and $s$ collectively as $q$.
The average lifetimes 
$\bar{\Gamma}_q \equiv (\Gamma_{qH}+\Gamma_{qL})/2$, 
mass differences 
$\Delta M_q \equiv M_{qH} - M_{qL}$,
lifetime differences 
$\Delta\Gamma_q \equiv \Gamma_{qL}-\Gamma_{qH}$, 
as well as CP asymmetries 
$\sin 2\beta_q$ with 
$\beta_q^{\rm SM} 
\equiv {\rm Arg}[-(V_{cb}^* V_{cq})/(V_{tb}^* V_{tq})]$
offer incisive probes of new physics.
Here the labels $L$ and $H$ stand respectively 
for the light and heavy mass eigenstates in the 
neutral $B_q$ system.
The values of $\Gamma_q, \Delta m_q$ and $\sin 2\beta_d$
have already been measured to an accuracy of better than 
$\sim 5\%$ \cite{pdg,deltams,utfit} and play an
important role in constraining any new physics.
The remaining quantities, on the other hand,
currently have large errors and their accurate
measurements act as tests of the SM. 
The value of $\sin 2\beta_s$, for example, which is
predicted to be $\approx -0.03$ in the SM, can
be enhanced significantly with many BSM physics
models \cite{betas-enhance} and a measurement in
excess of the SM prediction would vouch for the
presence of new physics.

The SM predicts the lifetime differences in the 
$B_d$ and $B_s$ system to be \cite{lenz-nierste}
$\dgd/\gd = (0.41^{+0.09}_{-0.10})\%$ and
$\dgs/\gs = (14.7 \pm 6.0) \%$ respectively,
$\dgd$ being suppressed with respect to $\dgs$
by a factor of $\sim |V_{td}/V_{ts}|^2 
\approx 0.05$.
The measurement of the latter is within the 
capability of LHCb, whereas that of the former
is very difficult even at the super-$B$ factories
due to its extremely small value.
The large theoretical uncertainties in the 
SM predictions for these two quantities make
them rather unsuitable for the detection
of BSM physics, unless such physics changes the value of 
$\Delta\Gamma_q/\Gamma_q$ beyond the SM uncertainties.
The so-called ``Grossman theorem'' 
states that new physics can only decrease the
value of $\dgs$ \cite{grossman}, and the result extended to
the $B_d$ system \cite{dhky} implies that 
$\dgd$ can increase at the most by 20\% 
with BSM contributions. 
This would seem to make the measurements of 
$\Delta\Gamma_q$ rather unappealing from the point
of view of detecting new physics.

However, the Grossman theorem, and its extension
mentioned above, are applicable only when
the BSM physics contributes to the dispersive part $M_{12q}$ of
the $\bqbqbar$ mixing amplitude, and not to its absorptive part
$\Gamma_{12q}$.
This is true for most of 
the BSM models, in particular the
minimal flavour violation (MFV) models \cite{mfv} 
where the CP violation emerges only from the CKM matrix. 
Even in non-MFV models, if the mixing box diagram contains
only heavy degrees of freedom, BSM physics cannot contribute
to $\Gamma_{12q}$. 
Such models include R-parity conserving supersymmetry, 
models with universal extra dimensions, little Higgs models, 
two-Higgs doublet models, etc.    
On the other hand, there are well-motivated models where 
the $\bqbqbar$
mixing box diagram contains two light degrees of freedom, 
resulting in an absorptive amplitude. We will discuss 
two such examples in this paper: 
(i) models with a scalar leptoquark, and 
(ii) R-parity violating (RPV) supersymmetry.
These models can have {\em flavour dependent} couplings of light known 
particles with one heavy new particle (squark or leptoquark), 
and hence can contribute to $\Gamma_{12q}$. 
This paves the way for an evasion of the Grossman bound, 
and potentially high lifetime differences
in both the $B_s$ and $B_d$ systems. 
We emphasize that these models are chosen just as examples 
and by no means
exhaust the list of all such possible models.

The new physics couplings also contribute to the mixing amplitudes 
(hence the mass splittings $\Delta M_q$ between the stationary 
states and the CP asymmetries), and to decay rates. 
As a consequence, the BSM 
parameter space is severely constrained by these data. 
In spite of these constraints, 
we show that these BSM models can indeed enhance the
lifetime difference in the $B_s$ up to its current
experimental limit, obtained from the angular analysis of
$B_s \to J/\psi \phi$ decays \cite{ddlr,ddf}.
In the $B_d$ system, the value of $\dgd/\gd$ can become as much 
as 2.5\% and hence come within the capabilities of the $B$ 
factories \cite{ckm-workshop}.
As a bonus, the mixing-induced CP asymmetry $\sin 2\beta_s$ 
can be enhanced by an order of magnitude from
its small SM value.
As an important byproduct, we also obtain limits on the
couplings of the BSM models considered above,
and predict enhanced branching ratios for decay channels
correlated with the enhanced lifetime differences.

The rest of the paper is organised as follows.
In Sec.~\ref{formalism}, we clarify the definitions,
methodology and approximations used to calculate the
SM as well as BSM contributions to $M_{12q}$ and 
$\Gamma_{12q}$.
In Sec.~\ref{numerical}, we present our numerical results, 
which give us limits on the new physics couplings,
as well as the enhancements of lifetime differences, 
$\beta_s$, and rates of correlated decay modes of $B_q$.
Sec.~\ref{summary} summarises our findings and 
discusses their implications for $B$ physics experiments.

\section{The formalism}
\label{formalism}

\subsection{The Standard Model}
\label{sm}

The effective Hamiltonian for the $\bqbqbar$ system in the 
flavour basis, with $CPT$ conservation, is given by
\be
H_{eff}=\begin{pmatrix}
M_{11q} -\textstyle{\frac{i}{2}} \Gamma_{11q} & 
M_{12q}-\textstyle{\frac{i}{2}}\Gamma_{12q} \cr
M_{12q}^\ast-\textstyle{\frac{i}{2}}\Gamma^\ast_{12q} &
M_{11q}-\textstyle{\frac{i}{2}}\Gamma_{11q} \end{pmatrix}.
\label{h-eff}
\ee
With the approximation $|\Gamma_{12q}| \ll |M_{12q}|$, 
which is valid for the $B_d$ as well as the $B_s$ system,
we have \cite{dhky}
\beq
\Delta M_q \approx 2 |M_{12q}| \; , \quad \quad
\Delta\Gamma_q \approx \frac{2 {\rm Re}(M_{12q} 
\Gamma_{12q}^*)}{|M_{12q}|} 
= 2 |\Gamma_{12q}| \cos \Phi_q \; ,
\label{dg-dgamma}
\eeq
where $\Phi_q$ is the phase difference between
$M_{12q}$ and $\Gamma_{12q}$.
This in fact demonstrates the Grossman theorem for $B_s$:
since $\Phi_s \approx 0$ in the SM, as long as there is
no BSM contribution to $\Gamma_{12s}$ itself, the value
of $\Delta\Gamma_s$ will always decrease due to the
cosine factor in (\ref{dg-dgamma}).

We now indicate the methodology of our calculations.
In the SM, the dispersive part of the $\bqbqbar$
mixing amplitude is given by \cite{hagelin}
\beq
M_{12q}^{\rm SM} = \frac{G_F^2}{12\pi^2} 
\hat{\eta}_{B_q} M_{B_q} B_{B_q} f_{B_q}^2 M_W^2 
(V_{tb}^* V_{tq})^2 S_0(x_t) \; ,
\label{m12q-sm}
\eeq
where $G_F$ is the Fermi constant, $M_X$ is the mass of
particle $X$, and $V_{ij}$s are the CKM matrix elements.
The short distance behaviour is contained in 
$\hat{\eta}_{B_q}$, which incorporates the QCD corrections,
and in the Inami-Lim function
\be
S_0(x) = \frac{4x-11x^2+x^3}{4(1-x)^2} - \frac{3x^3\ln x}
{2(1-x)^3} \; ,
\label{inami}
\ee
with $x_f$ for a fermion $f$ defined by
\beq
x_f \equiv m_f^2/M_W^2 \; .
\label{xf-def}
\eeq
The decay constant $f_{B_q}$ and the bag factor $B_{B_q}$
take care of the hadronic matrix element 
\beq
\bra \bar{B}_q |(\bar{b}q)_{V-A} (\bar{b}q)_{V-A} | B_q \ket
= (8/3) M_{B_q}^2 f_{B_q}^2 B_{B_q} \; ,
\label{q-matrix}
\eeq
with the wavefunction for the $B_q$ meson normalised as
\beq
2M_{B_q}|M_{12q}| = \bra \bar{B_q} | H_{eff} | B_q\ket \; .
\eeq

The absorptive part of the $\bqbqbar$ mixing amplitude
in the SM, to leading order in QCD, is given by
\cite{hagelin,gamma12-lo}
\be
\Gamma_{12q}^{\rm SM (0)} 
= - \frac{G_F^2f_{B_q}^2 B_{B_q}M_{B_q}}{8\pi} 
(V_{cb}^* V_{cq})^2 m_b^2 F(c) \; ,
\ee
where
\be
F(f) = \sqrt{1-4\frac{m_f^2}{m_b^2} } \left( 1- \frac{2}{3}
\frac{m_f^2}{m_b^2} \right) \; 
\label{f-def}
\ee
for a fermion $f$.
The next to leading order QCD corrections and the
$1/m_b$ corrections modify the value of $\Gamma_{12s}$ 
\cite{nlo-dgs} as well as $\Gamma_{12d}$ \cite{dhky}
in the SM by $\sim 30\%$.
The current theoretical status of $\dgq/\gq$ has been
summarised in \cite{lenz-nierste}. 

In the presence of any BSM contribution, we have
\beq
M_{12q} = M_{12q}^{\rm SM} + M_{12q}^{\rm BSM} \; , \quad \quad
\Gamma_{12q} = \Gamma_{12q}^{\rm SM} + 
\Gamma_{12q}^{\rm BSM} \; .
\eeq
In our numerical analysis in Sec.~\ref{numerical}, we use
the SM predictions \cite{lenz-nierste}
that include the NLO QCD and $1/m_b$ corrections,
however for $\Gamma_{12q}^{\rm BSM}$ 
we only use the leading order contributions 
$\Gamma_{12q}^{\rm BSM(0)}$.
Note that the QCD corrections are expected to be different for SM
and BSM operators since the mediating heavy particle for the latter
case is a colour triplet. The $1/m_b$ corrections are also expected
to differ since the light degrees of freedom that flow inside the
mixing box are different too. While it is desirable to have an
idea of these corrections, in this work we will just assume that
these corrections are likely to introduce an error of $\sim 30\%$
in our calculations. 
Since our final results claim enhancements of
up to 5 times over the lifetime differences 
in the SM, and the BSM model calculations themselves
depend on the unknown masses of the new particles,
the higher order corrections would not change the 
conclusions qualitatively.

\subsection{Leptoquark Models}
\label{leptoquark}

Leptoquarks are colour-triplet objects that couple to
quarks and leptons. They occur generically in 
GUTs \cite{lq-gut}, composite models \cite{lq-composite},
and superstring-inspired $E_6$ models \cite{lq-e6}.
Model-independent constraints on their properties are
available \cite{davidson-zpc61}, and the prospects of
their discovery at the LHC have also been studied \cite{mitsou}.
We shall restrict ourselves to scalar leptoquarks that are
singlets under the $SU(2)_L$ gauge group of the SM.
This is because vector or $SU(2)_L$ nonsinglet leptoquarks tend to 
couple directly to neutrinos, hence we expect that 
their couplings are tightly constrained from the neutrino mass and
mixing data.
This makes any significant effect on the $\bqbqbar$ system
unlikely.

The relevant interaction term for a scalar leptoquark 
$S_0$ is of the form
\be
{\cal L}_{\rm LQ} = \lambda_{ij} \bar{d^c}_{jR} 
e_{iR}S_0 + {\rm h.c.}  \; ,
\label{lqlag}
\ee
where $d_R$ and $e_R$ stand for the right-handed down-type
quarks and right-handed charged leptons respectively, and
$i,j$ are generation indices that run from 1 to 3.
The couplings $\lambda_{ij}$ can in general be complex,
and some of them may vanish depending on any flavour
symmetries involved.

When $\lambda_{32}$ and $\lambda_{33}$ are nonzero,
the interaction (\ref{lqlag}) generates
an effective four-fermion
$(S+P)\otimes (S+P)$ interaction leading to $b\to s\tau^+\tau^-$. 
This will contribute to $\bsbsbar$ mixing 
(with $\tau$ and $S^0$ flowing inside the box), 
to the leptonic decay $B_s\to \tau^+\tau^-$, and to the semileptonic
decays $B_q\to X_s \tau^+\tau^-$.
The relevant quantity here is the coupling product
\beq
h_{LQ} (b \to s \tau^+ \tau^-) \equiv 
\lambda_{32}^\ast \lambda_{33} \; ,
\eeq
such that $h_{LQ} (b \to s \tau^+ \tau^-)/(8 M_{S_0}^2)$ is
the effective leptoquark coupling equivalent to 
$(G_F/\sqrt{2}) V_{tq}^\ast V_{tb}$ or 
$(G_F/\sqrt{2}) V_{cq}^\ast V_{cb}$ 
in the SM.

By changing the leptonic index from 3 to 2,
one gets the second generation leptoquark 
that can lead to $b \to s \mu^+ \mu^-$ with 
the relevant coupling product
\beq
h_{LQ} (b \to s \mu^+ \mu^-) \equiv 
\lambda_{22}^\ast\lambda_{23} \; .
\eeq
In addition to $\bsbsbar$ mixing, the coupling 
$h_{LQ}(b \to s \mu^+ \mu^-)$ contributes also to
$B_s \to \mu^+ \mu^-$ and $B_q \to X_s \mu^+ \mu^-$.
The upper bound on the branching ratio of $B_s \to \mu^+ \mu^-$
constrains this coupling product severely, as will be seen in 
Sec.~\ref{numerical}.
One can also have mixed leptonic indices, giving rise
to the channel $b \to s \tau^+ \mu^-$ for example, which we
will not discuss here. 
One expects the first generation leptoquarks to be heavier, from
the Tevatron and HERA data \cite{bounds-gen1-lq}, so 
the  couplings of the first generation leptoquarks 
are highly constrained, and we do not consider them in this paper.
However, note that the bounds are, in general, dependent on the 
quantum numbers of the leptoquarks.

The $\bdbdbar$ mixing is modified by the 
coupling product
\beq
h_{LQ} (b \to d \tau^+ \tau^-) = \lambda_{31}^\ast\lambda_{33}
\; , \quad \quad
h_{LQ} (b \to d \mu^+ \mu^-) = \lambda_{21}^\ast\lambda_{23} \; .
\eeq
The former combination $h_{LQ} (b \to d \tau^+ \tau^-)$ affects
$B_d \to \tau^+ \tau^-$ and $B_d \to X_d \tau^+ \tau^-$,
whereas the latter affects 
$B_d \to \mu^+ \mu^-$ and $B_d \to X_d \mu^+ \mu^-$.

In terms of the coupling products $h_{LQ}$, the contributions of
the leptoquarks to the $\bqbqbar$ mixing is given by
\bea
M_{12q}^{\rm LQ} & = & \sum_{\ell = \mu, \tau} 
\frac{h_{LQ}^2(b \to q \ell^+ \ell^-)}{384\pi^2 M_{S_0}^2} 
M_{B_q} \hat{\eta}_{B_q} f_{B_q}^2 B_{B_q} 
\tilde{S_0}(x_\ell) \; ,\nonumber\\
\Gamma_{12q}^{\rm LQ(0)} & = & - \sum_{\ell = \mu, \tau} 
\frac{h_{LQ}^2(b \to q \ell^+ \ell^-)}{256 \pi M_{S_0}^4} 
M_{B_q} f_{B_q}^2 B_{B_q} m_b^2 F(\ell) \; .
\label{mgamma-lq}
\eea
where the function $\tilde{S_0}(x)$ is
\be
\tilde{S_0}(x) = \frac{1+x}{(1-x)^2} + \frac{2x\log x}{(1-x)^3}\; ,
\label{s0tilde}
\ee
and $F(f)$ is as given in (\ref{f-def}).
While calculating the limits on the new physics parameter space,
we assume that at a time, only one of 
$h_{LQ}(b \to q \tau^+ \tau^-)$ and 
$h_{LQ}(b \to q \mu^+ \mu^-)$ is nonvanishing,
so that the right hand side of each of the equation in 
(\ref{mgamma-lq})
has only one term.

As noted above, leptoquarks contribute to the leptonic
decay $B_q \to \ell^+ \ell^-$.
In the SM, this decay rate is extremely small \cite{bqll-sm}:
$BR(B_d\to \mu^+\mu^-)\approx 1.1\times 10^{-10}$, 
$BR(B_d\to\tau^+\tau^-) \approx 3.1\times 10^{-8}$.
We therefore approximate the branching fraction of
$B_q \to \ell^+ \ell^-$ by only the
leptoquark contribution \cite{gln-tau}: 
\be
{\rm BR}(B_q\to\ell^+\ell^-) \approx 
\frac{|h_{LQ}(b \to q \ell^+ \ell^-)|^2}{ 128 \pi M_{S_0}^4}
 \frac{f_{B_q}^2 M_{B_q}^3}{\Gamma_{B_q}} \frac{m_\ell^2}
{M_{B_q}^2} \sqrt{1-4\frac{m_{\ell}^2}{M_{B_q}^2}} \; .
\ee
The presence of leptoquarks may be detected through the
measurement of such an enhanced branching fraction.
On the other hand, upper bounds on these branching
fractions constrain the leptoquark coupling products
discussed above.

\subsection{R-parity violating supersymmetry (RPV SUSY)}
\label{rpv-susy}

R-parity is the discrete symmetry defined as $R=(-1)^{3B+L+2S}$, where 
$B$, $L$, and $S$ are respectively baryon number, lepton number and spin 
of a particle. $R$ equals $1$ for all SM particles and $-1$ for
all superparticles. 
Though R-parity is conserved {\em ad hoc}
in a large class of supersymmetric models,
one can write R-parity violating terms
respecting Lorentz and gauge invariance. These terms can violate either 
$B$ or $L$, but not both, since that will lead to uncontrollably large
proton decay rate. 
In this work, we consider only the $B$-violating terms.
Most of the $L$-violating $\lambda'$-type couplings 
are highly constrained from neutrino mass \cite{dreiner}
and $\Delta M_s$ measurement \cite{nandi-saha}, 
so their contribution to the neutral meson mixing 
is expected to be small.

Consider the $B$ violating term in the superpotential
\be
{\cal W}=\epsilon_{\alpha\beta\gamma}\lambda''_{ijk} U^c_{i\alpha} D^c_{j
\beta} D^c_{k\gamma} \; ,
\label{superpot}
\ee
where $\alpha,\beta,\gamma$ are colour indices. 
The couplings $\lambda''_{ijk}$ are in general complex,
and are antisymmetric
in the last two generation indices $j$ and $k$, 
{\em i.e.}, $\lambda''_{ijk}=-\lambda''_{ikj}$. 
In terms of the component fields, this can give rise to terms
in the Lagrangian that are of the form
\be
\epsilon_{\alpha\beta\gamma}\lambda''_{ijk} \tilde{u}^c_{i\alpha} 
\bar{d}_{j\beta} P_R d^c_{k\gamma} \; , \quad  {\rm or} \quad 
\epsilon_{\alpha\beta\gamma}\lambda''_{ijk} \bar{u}_{i\alpha} 
P_R d^c_{k\gamma} \tilde{d}^c_{j\beta} \; , 
\label{rpv-lag}
\ee
where $P_R=(1+\gamma_5)/2$. As $j\not = k$, the only possible combinations
responsible for $\bsbsbar$ mixing are $\lambda''_{i13} 
\lambda''^{\ \ast}_{i12}$.
For $\bdbdbar$ mixing, the relevant combination is
$\lambda''_{i23} \lambda''^{\ \ast}_{i21}$.
As usual, we consider only one product to be nonzero at a time. This also
helps us avoid the tighter constraints coming from, say, 
$K^0-\bar{K}{}^0$ 
mixing.

\begin{figure}
\epsfig{file=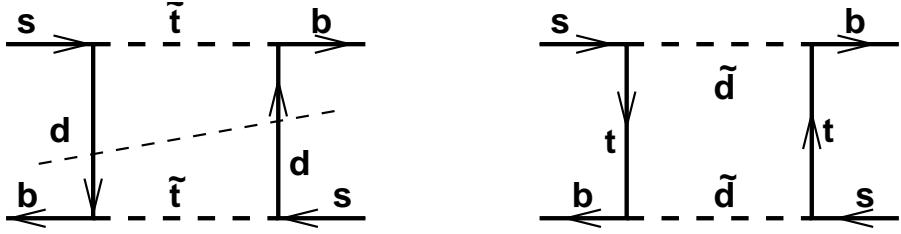,width=12cm}
\caption{$\bsbsbar$ mixing diagrams with RPV SUSY. Note that only the
first diagram gives an absorptive part. To get the total amplitude, one
must also include the crossed boxes.
\label{feynman}}
\end{figure}

Let us first consider $\bsbsbar$ mixing. 
The $\lambda''$ couplings with $i=1$ are
highly constrained to be at most $\sim {\cal O}(10^{-4}-10^{-5})$
from $n$-$\bar{n}$ oscillation and double nucleon decay
\cite{rpv-gen1-couplings}. 
Though there is no significant bound on the $i=2$ couplings, 
the corresponding coupling products, 
if comparable with the SM, would contribute and modify
$b\to c\bar{c}s$ processes. Just to avoid this consequence, we do not
consider the $i=2$ couplings any further and move to nonzero values of
$i=3$ couplings.
This leads to two new mixing diagrams as shown in Fig.~\ref{feynman}: 
one with internal $d$ quarks and $\tilde t$ squarks,
and the other with internal $t$ quarks and $\tilde d$ squarks. 
While both these diagrams contribute to $M_{12s}$, 
only the former has an absorptive contribution that goes
towards $\Gamma_{12s}$. This leads to
\bea
M_{12s}^{\rm RPV} & = & \frac{h_{\rm RPV}^2(b \to s)}{192 \pi^2 
M_{\tilde q_R}^2}  M_{B_s} \hat{\eta}_{B_s}
f_{B_s}^2 B_{B_s} 
\left(\tilde{S_0}(x_t) + \tilde{S_0}(x_d)\right),\nonumber\\ 
\Gamma_{12s}^{\rm RPV(0)} & = & -\frac{h_{\rm RPV}^2(b \to s)}{128 \pi
M_{\tilde q_R}^4} 
M_{B_s}f_{B_s}^2 B_{B_s}m_b^2 F(d) \; ,
\label{mgamma-s-rpv}
\eea
where 
\beq
h_{\rm RPV}(b \to s) \equiv
\lambda''^{\ \ast}_{313} \lambda''_{312} \; 
\eeq
is the relevant coupling product in RPV SUSY, 
while the factors $\tilde{S_0}(x)$ and $F(d/s)$ are as given in 
(\ref{s0tilde}) and (\ref{f-def}) respectively.
We have assumed the relevant squarks, $\tilde{t}_R$ and $\tilde{d}_R$, 
to be degenerate in mass.

The coupling product $h_{\rm RPV}(b \to s)$ also
contributes to nonleptonic $B$ decays taking place via
$b \to d \bar{d} s$, like
$B^+ \to K^0 \pi^+, B^+ \to K^+ \pi^0, B_d \to K^0 \pi^0,
B_s \to \phi \pi^0, B_s \to \pi^+ \pi^-$ and their CP conjugate decays.
The decay rates depend on the coherent sum of the SM and the BSM
amplitudes. The former is calculated in the naive factorisation
model \cite{ali-kramer-lu}. However, considering the uncertainties
in any such calculation, we have been slightly conservative: the form
factors are not directly taken from \cite{ali-kramer-lu} but fitted 
so that even without the BSM part, the pure SM expectation of the
branching ratio is consistent with the data. This means that the
branching ratio data may not be used for claiming the presence of
BSM physics; it can, at best, constrain the parameter space from above.
The BSM part is computed with some further simplifying assumptions. 
We assume the strong phase difference to be zero between the SM and the
BSM amplitudes. However, since the weak phase of the BSM coupling is varied
over its full range, this effect is offset for $B^+$ decays. We have also
not taken into account the mixing between the RPV operators and the SM
operators between the scales $M_W$ and $m_b$. The dominant effect, which
is just a multiplicative renormalisation of the RPV operator, can be taken
into account by redefining the couplings so that their values are calculated
at the scale $m_b$ and not at the high scale (thus, one should be careful
in using the constraints on the couplings; to get the constraints at a
higher scale, one must run them upwards).

The $\bdbdbar$ system may be analyzed in an analogous manner,
simply with the interchange $d \leftrightarrow s$.
With the naive factorization approximation, The mixing amplitude 
has the dispersive and the absorptive parts
\bea
M_{12d}^{\rm RPV} & = & \frac{h_{\rm RPV}^2(b \to d)}{192 \pi^2 
M_{\tilde q_R}^2}  M_{B_d} \hat{\eta}_{B_d}
f_{B_d}^2 B_{B_d} 
\left(\tilde{S_0}(x_t) + \tilde{S_0}(x_s)\right),\nonumber\\ 
\Gamma_{12d}^{\rm RPV(0)} & = &-\frac{h_{\rm RPV}^2(b \to d)}{128\pi
 M_{\tilde q_R}^4}
M_{B_d}f_{B_d}^2 B_{B_d}m_b^2 F(s) \; ,
\eea
where the coupling product involved is
\beq
h_{\rm RPV}(b \to d) \equiv
\lambda''^{\ \ast}_{323} \lambda''_{321} \; ,
\eeq
and the notation used is the same as in the $\bsbsbar$ case.
One may note that in a number of well-motivated models, the lighter
stop may be significantly lighter than the other squarks. In this case,
the value of $\dgs$ ($\dgd$) would be higher than the one computed here.

The coupling product $h_{\rm RPV}(b \to d)$ also 
contributes to the nonleptonic $b \to s \bar{s} d$ decay
channels like $B^{+,0}\to\pi^{+,0}\phi$.
Indeed, the measurement of BR$(B^+ \to \phi \pi^+)$ provides
the strongest constraint on $|h_{\rm RPV} (b \to d)|$.

Note that the relations for RPV SUSY contain a relative factor 
of two compared to the leptoquark case; this is due to the fact 
that in RPV, the effective $|\Delta B| = 1$ Hamiltonian 
contribution to $b \to d \bar{d} s$ ($b \to s \bar{s} d$) 
in the $B_s$ ($B_d$) system contains two terms,
which come from the contraction of two colour 
$\epsilon$-factors in (\ref{rpv-lag}): 
$\epsilon_{abc}\epsilon_{ade}=\delta_{bd}
\delta_{ce}-\delta_{be}\delta_{cd}$ where $\delta$ is the 
Kronecker delta function.

\section{Numerical results}
\label{numerical}

In this section we shall numerically evaluate 
the allowed values of 
$\Delta\Gamma_q/\Gamma_q$ and $\sin 2\beta_s$ in the
framework of the BSM models considered in the previous 
section, applying the constraints from current
measurements. Unless otherwise mentioned, all numbers are taken from
\cite{hfag}.
The average lifetimes of $B_q$ mesons,
\beq
\Gamma_d = (0.653 \pm 0.004)~{\rm ps}^{-1} \; , \quad \quad 
\Gamma_s = (0.682 \pm 0.027)~{\rm ps}^{-1} \; ,
\eeq
are not affected much by the BSM physics.
The measured values of the mass differences are
\be
\delmd = (0.507\pm 0.004)~{\rm ps}^{-1} \; , \quad \quad
\delms = (17.33^{+0.42}_{-0.21} \pm 0.07) ~{\rm ps}^{-1} \; .
\label{deltams}
\ee
The value of $\sin 2\beta_d$ in the SM, obtained from a
fit that does not involve $\bdbdbar$ mixing, is \cite{utfit}
\beq
\sin 2\beta_d^{\rm SM} = 0.752 \pm 0.040
\eeq
whereas the CP asymmetry in the $\bdbdbar$ system
is measured, from charmonium modes, to be \cite{utfit}
\be
\sin 2\beta_d =0.674\pm 0.026 \; .
\label{betad}
\ee
We constrain the BSM physics by requiring $\sin 2\beta_d$
to lie in this interval.
The current limits on the values of $\Delta\Gamma_q$
and $\beta_s$ \cite{utfit,d0-dgammas,hfag},
\be
\frac{\delgs}{\Gamma_s} = 0.206^{+0.106}_{-0.111}\; , \quad 
\frac{\delgd}{\Gamma_d} = 0.009 \pm 0.037 \; , \quad
\beta_s = -0.79 \pm 0.56 \; ,
\label{dgs-lim}
\ee
though very weak, are also included for consistency.
In particular, it will be seen that in a model with third
generation leptoquark, the possible value of $\dgs/\gs$ , which 
should lie in $(-0.01,0.51)$ at 95\% C.L. 
\footnote{The likelihood distribution of $\dgs/\gs$ is
extremely skewed \cite{hfag}, 
so that the upper limit of the confidence
interval is rather high compared to the naive 2$\sigma$ 
estimate. Note that we have used the 
Heavy Flavor Averaging Group (HFAG) limits
obtained from a fit to only the direct measurements of the 
lifetime difference. If combined with the lifetimes obtained
from flavour specific modes and $B_s \to K^+ K^-$, the 
95\% C.L. upper
bound on $\dgs/\gs$ will decrease to 0.25 \cite{hfag}.}
\cite{hfag}, is bounded only by the experimental limit.

We also use the values of the bag factors 
\cite{utfit,ckmfitter}
\be
f_{B_d} \sqrt{B_{B_d}}= 223\pm 35~{\rm MeV} \; , \quad \quad
f_{B_s} \sqrt{B_{B_s}}= 262\pm 35~{\rm MeV} \; ,
\eeq
and the short distance factors
\beq
\hat{\eta}_{B_d} = \eta_{B_s} =0.55 \; , \quad \quad 
S_0(x_t) = 2.327 \pm 0.044 \; .
\ee
The relevant CKM elements are
\be
|V_{td}|=8.54(28)\times 10^{-3} \; ,\quad \quad 
|V_{ts}|=40.96(61)\times 10^{-3} \; ,
\ee
while the other elements are taken to be fixed at their central 
values.
For leptonic decays, we use $f_{B_s} = f_{B_d} = 200$ MeV.

To constrain the BSM effects, we present our results taking the SM 
contribution to $\dgq$ to be $\delgs= (0.096 \pm 0.039)$ ps$^{-1}$
and $\delgd=(26.7 \pm 6.1) \times 10^{-4}$ ps$^{-1}$ \cite{lenz-nierste}. 
The SM theoretical uncertainty is about 30\% for $\delgs$ and 
about 25\% for $\delgd$, so one must be careful
in interpreting the results. 
Only when the values of $\dgq$ in the presence of BSM 
physics exceed the SM calculations beyond their upper
limit including the large error bars can the presence of
new physics be claimed. As will be seen in the following,
it is indeed possible in the scalar leptoquark model with a
third generation leptoquark.

The new physics models are parametrised by the magnitude 
of the relevant coupling product $h_{\rm BSM}$ and its weak phase. 
We vary these two as free parameters and scan the 
parameter space, 
taking the SM parameters to have Gaussian distributions with
the means and standard deviations as given above.
We require the calculated quantities to be within the 
95\% C.L. (2$\sigma$) intervals of the experimental
measurements, where such a measurement is available
(e.g. $\Delta M_q$, $\Delta \Gamma_q$). 
Whenever only an upper bound is available (e.g.
some branching ratios), we require
the calculated quantities to be within the 90\% C.L.
interval, since these are the quoted limits \cite{hfag}.

\subsection{Leptoquark}
\label{leptoquark-numerical}

As pointed out in Sec.~\ref{leptoquark}, the couplings that
may enhance $\dgq/\gq$ also tend to enhance the branching
ratio $B_q \to \ell^+ \ell^-$, where $\ell$ is the charged
lepton of the corresponding leptoquark generation.
The limits coming from the measurements of these decay modes
therefore constrain the allowed ranges of BSM parameters.
The relevant branching ratios are \cite{hfag}
\bea
{\rm BR}(B_d \to \mu^+\mu^-) < 2.3\times 10^{-8} \; , 
& \quad \quad & 
{\rm BR}(B_d \to \tau^+\tau^-) < 4.1\times 10^{-3} \; , 
\ \ \nonumber \\ 
{\rm BR}(B_s \to \mu^+\mu^-) < 7.5 \times 10^{-8} \; . & & \; 
\eea
The leptoquark mass is taken to be 100 GeV.

The severe constraints on $B_q \to \mu^+ \mu^-$ restricts the
coupling products of the second generation leptoquark to 
$|h_{LQ}(b \to d\mu^+ \mu^-)| < 4 \times 10^{-4}$ 
and $|h_{LQ}(b \to s\mu^+ \mu^-)| < 6 \times 10^{-4}$, 
as a consequence there is no significant enhancement in either
$\dgq/\gq$ or $\beta_s$. 
Since the couplings of the third
generation leptoquark are not so severely restricted,
it is possible to enhance the values of these quantities. 

We show in Fig.~\ref{lq3-bs} our results in the $B_s$ system
with the third generation leptoquark.
We display the allowed values of the phase of the coupling product, 
${\rm Arg}[h_{LQ}(b \to s \tau^+ \tau^-)]$,
as well as the allowed values of $\dgs/\gs$, 
$\beta_s$ and ${\rm BR}(B_s \to \tau^+ \tau^-)$,
as functions of the magnitude of the coupling product,
$|h_{LQ}(b \to s \tau^+ \tau^-)|$.
We observe the following:

\begin{figure}[htbp]
\parbox{8cm}{
\epsfig{file=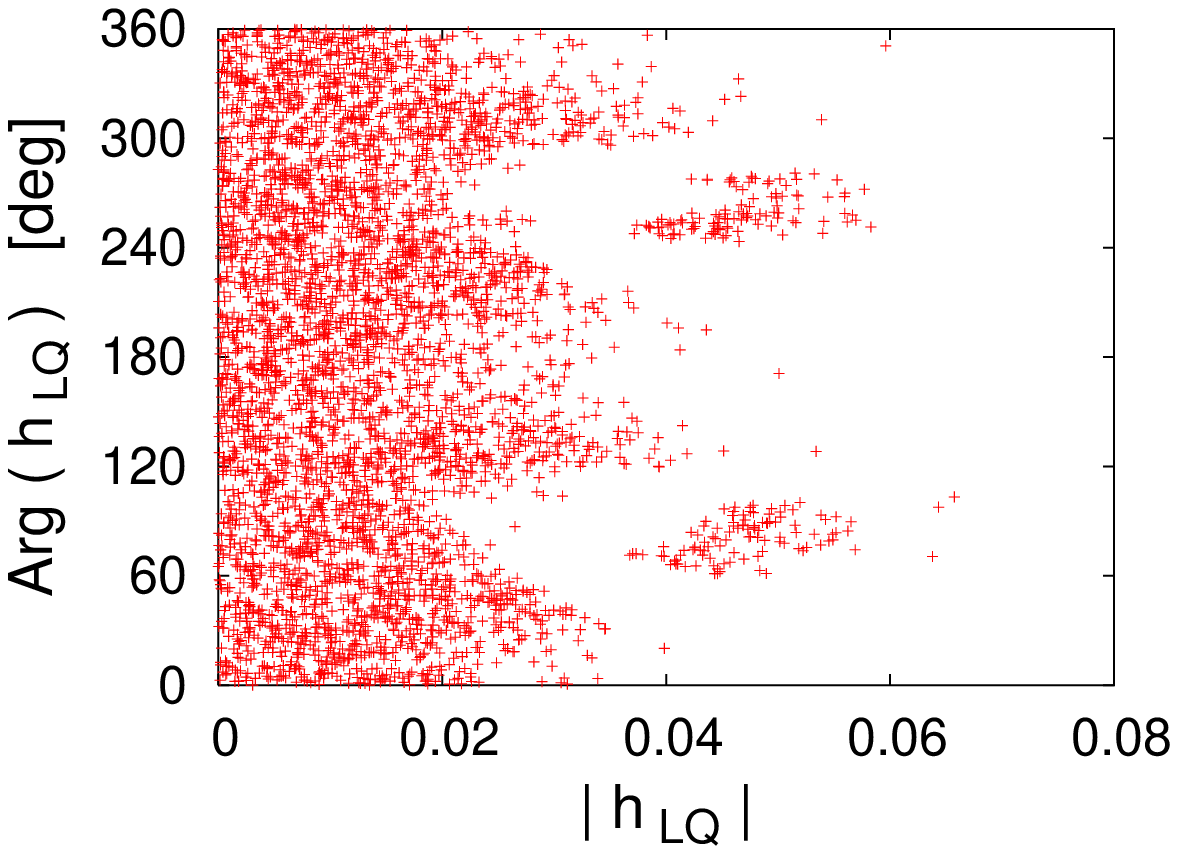,width=8cm,angle=0}
}
\parbox{8cm}{
\epsfig{file=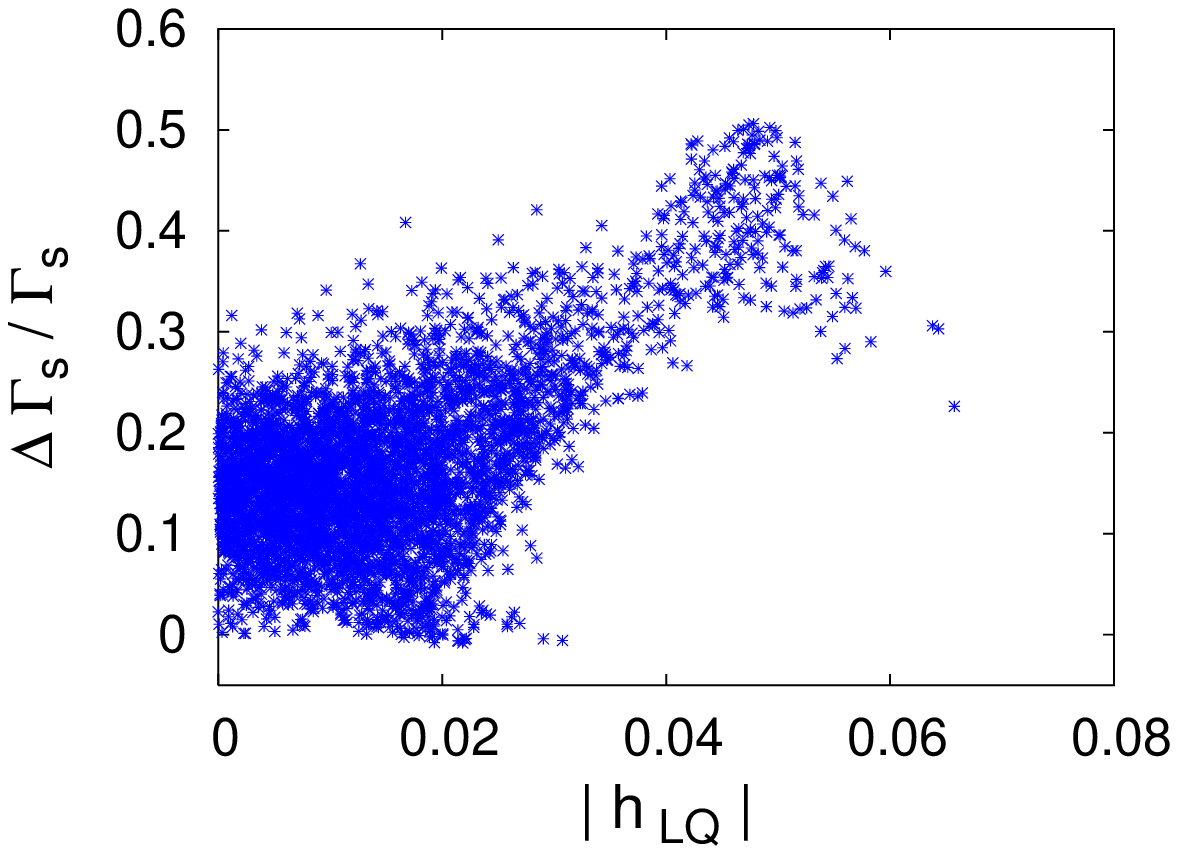,width=8cm,angle=0}
}

\parbox{8cm}{
\epsfig{file=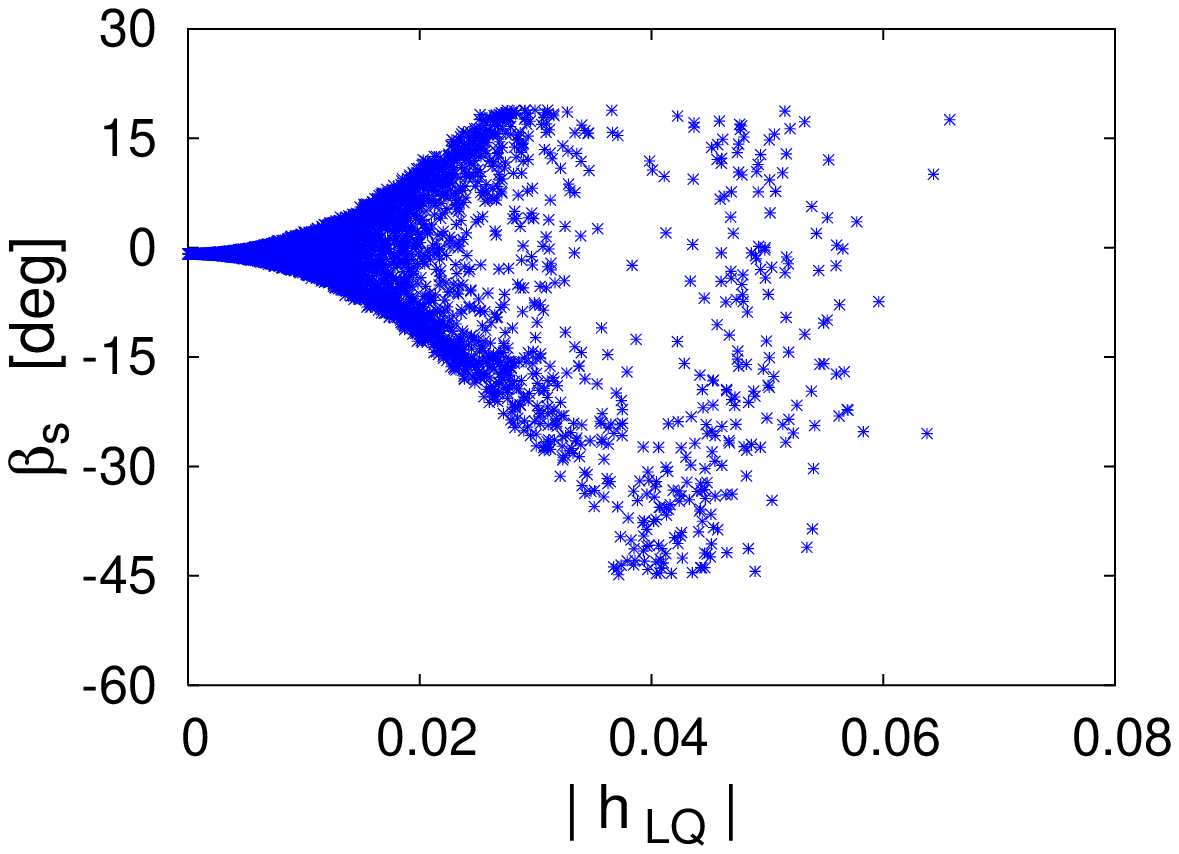,width=8cm,angle=0}
}
\parbox{8cm}{
\epsfig{file=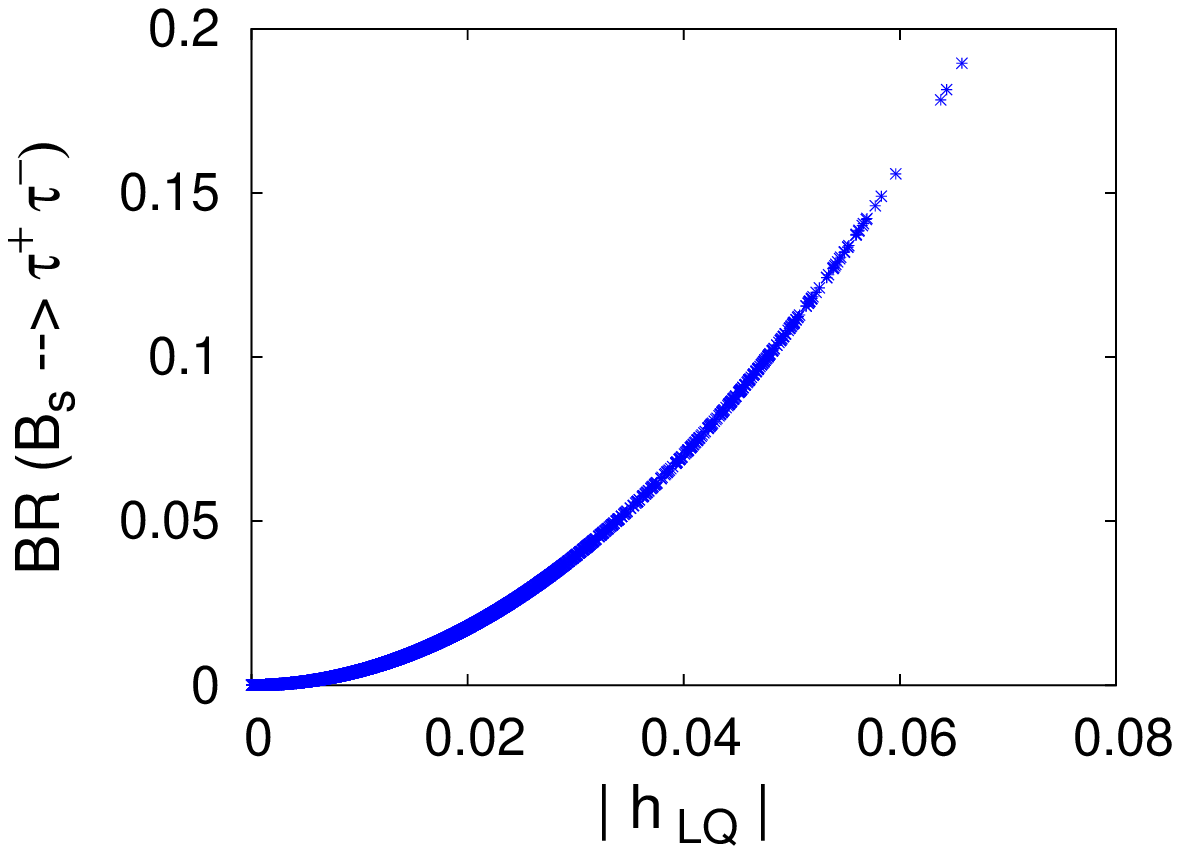,width=8cm,angle=0}
}
\caption{
The allowed values of the phase of the coupling product, 
${\rm Arg}[h_{LQ}(b \to s \tau^+ \tau^-)]$,
as well as the corresponding allowed values of
$\dgs/\gs$, $\beta_s$ and ${\rm BR}(B_s \to \tau^+ \tau^-)$
as functions of the magnitude of the coupling product,
$|h_{LQ}(b \to s \tau^+ \tau^-)|$.
The mass of the leptoquark is taken to be 100 GeV.
Note that we have used the 95\% C.L. bounds on $\dgs/\gs$
from only its direct measurements. 
\label{lq3-bs}}
\end{figure}

\begin{itemize}

\item
The major constraints on the parameter space of
$h_{LQ}$ arise from
the $\Delta M_s$ measurement (\ref{deltams}).
The allowed parameter space is not continuous,
but has small islands at higher values of $|h_{LQ}|$.
This is due to the constructive (destructive) interference 
between the SM and the BSM amplitudes for $\Delta M_s$,
which forbids (allows) some values of ${\rm Arg}(h_{LQ})$
with the $\Delta M_s$ measurements.

\item 
The value of $\dgs/\gs$ is bounded from above only by the current 
experimental limit (\ref{dgs-lim}), which is more than three times the 
central value predicted by the SM, and more than $5\sigma$
away from it even when the theoretical uncertainties
are included. If indeed a third generation leptoquark of
mass $\sim 100$ GeV is present, the measurement of $\dgs/\gs$
is literally round the corner, perhaps even possible at
the Tevatron \cite{durham}.

\item 
The value of $\beta_s$ can become as high as $20^\circ$ or 
as low as $-45^\circ$, which makes its measurement with the
decay modes like $B_s \to J/\psi \eta^{(')}$,
or through the angular distributions of
$B_s \to J/\psi \phi$ \cite{ddlr} or
$B_s\to D_s^{+(*)}D_s^{-(*)}$ \cite{ddf},
easily feasible 
at the $B$ factories or at the LHC experiments 
\cite{ball-lhc} .

\item
A large value of $\dgs/\gs$ is also accompanied by a 
large branching ratio $B_s \to \tau^+ \tau^-$,
which may be enhanced to as much as 18\%.
Currently no measurement of this decay channel is
available, however if it indeed has such a large
decay rate, its measurement would also indicate a large
value of $\dgs$.

\end{itemize}

\begin{figure}[htbp]
\parbox{8cm}{
\epsfig{file=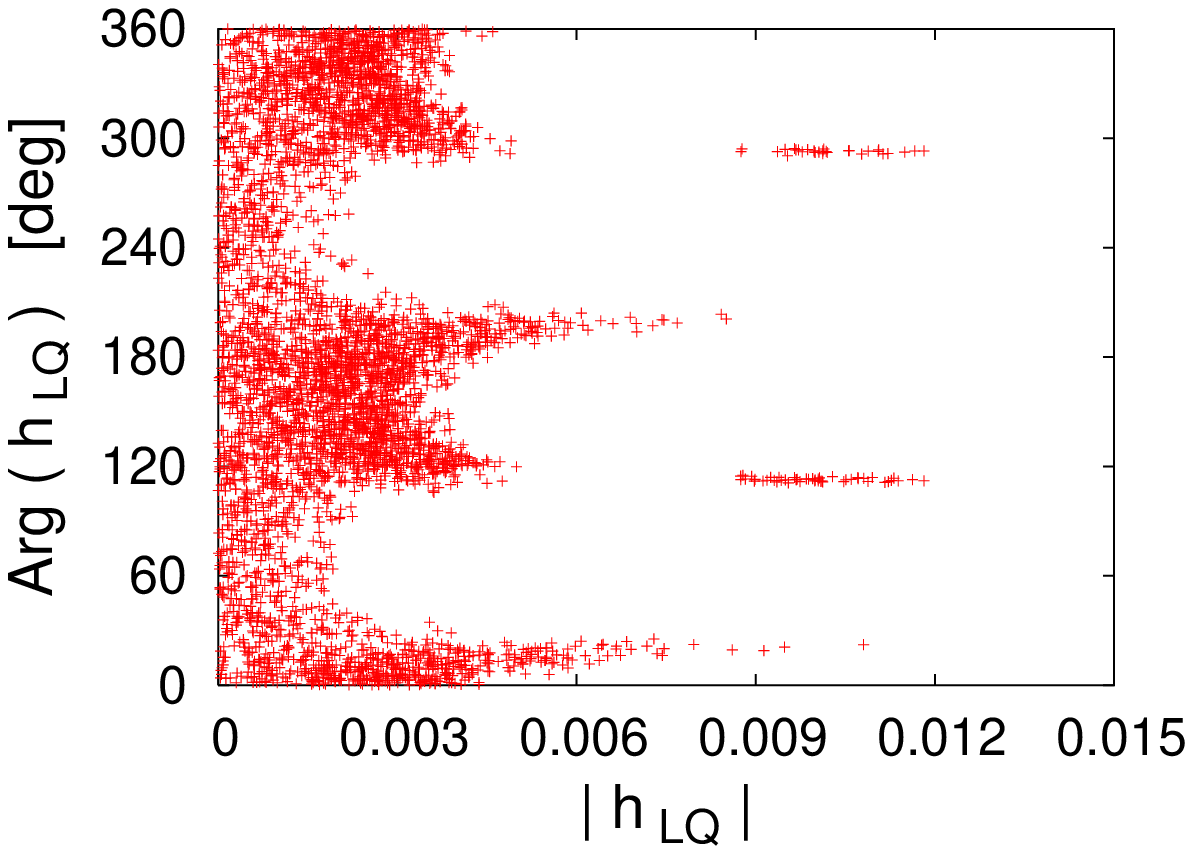,width=8cm,angle=0}
}
\parbox{8cm}{
\epsfig{file=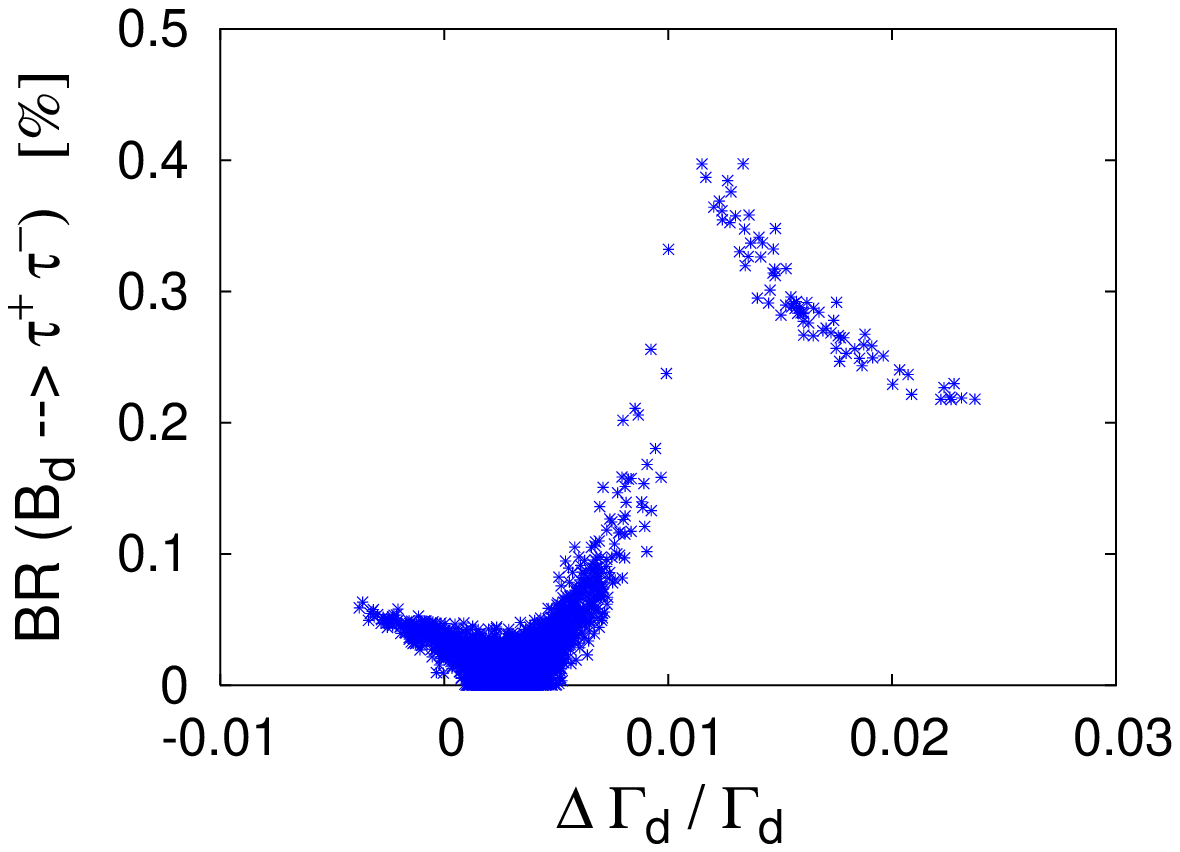,width=8cm,angle=0}
}
\caption{
The left panel shows the allowed parameter space of the 
phase and magnitude of the coupling product
$h_{LQ}(b \to d \tau^+ \tau^-)$.
The right panel shows the allowed values of $\dgd/\gd$
along with their correlation with ${\rm BR}(B_d \to \tau^+
\tau^-)$.
The mass of the leptoquark has been taken to be 100 GeV.
\label{lq3-bd}}
\end{figure}

Analogous results with the $\bdbdbar$ system are
displayed in Fig.~\ref{lq3-bd},
where we show the allowed parameter space of the
magnitude and phase of $h_{LQ}(b \to d \ell^+ \ell^-)$,
as well as the corresponding allowed 
$\dgd/\gd$ values and their correlations with
${\rm BR}(B_d \to \tau^+ \tau^-)$. 
Here, the mass difference $\Delta M_d$ 
as well as the well measured value of $\sin 2\beta_d$ 
(\ref{betad}) restrict the leptoquark coupling. 
It may be seen from the figure that:
\begin{itemize}
\item A rather specific
value of the relative phase between the SM and the 
BSM amplitudes allows one to obey the experimental
constraints while at the same time allowing for the
value of $\dgd/\gd$ as high as 2.5\%. 
\item
Such a high lifetime difference also 
comes in conjunction with a  $B_d \to \tau^+ \tau^-$
branching ratio as high as $0.4 \%$, which is just 
below the current experimental upper bound. 

\end{itemize}

Thus, the mediation due to a third generation scalar leptoquark
is able to increase $\dgd/\gd$ to a value large enough to be
measurable at the $B$ factories \cite{ckm-workshop}. 
This should be a strong motivation
for trying to measure this quantity in the $B_d$ system.
At the same time, the measurement of 
${\rm BR}(B_d \to \tau^+ \tau^-)$ should also be able to 
indicate whether $\dgd/\gd$ is significant or not.

\subsection{R-parity violating supersymmetry}
\label{rpv-susy-numerical}

\begin{figure}[htbp]
\parbox{8cm}{
\epsfig{file=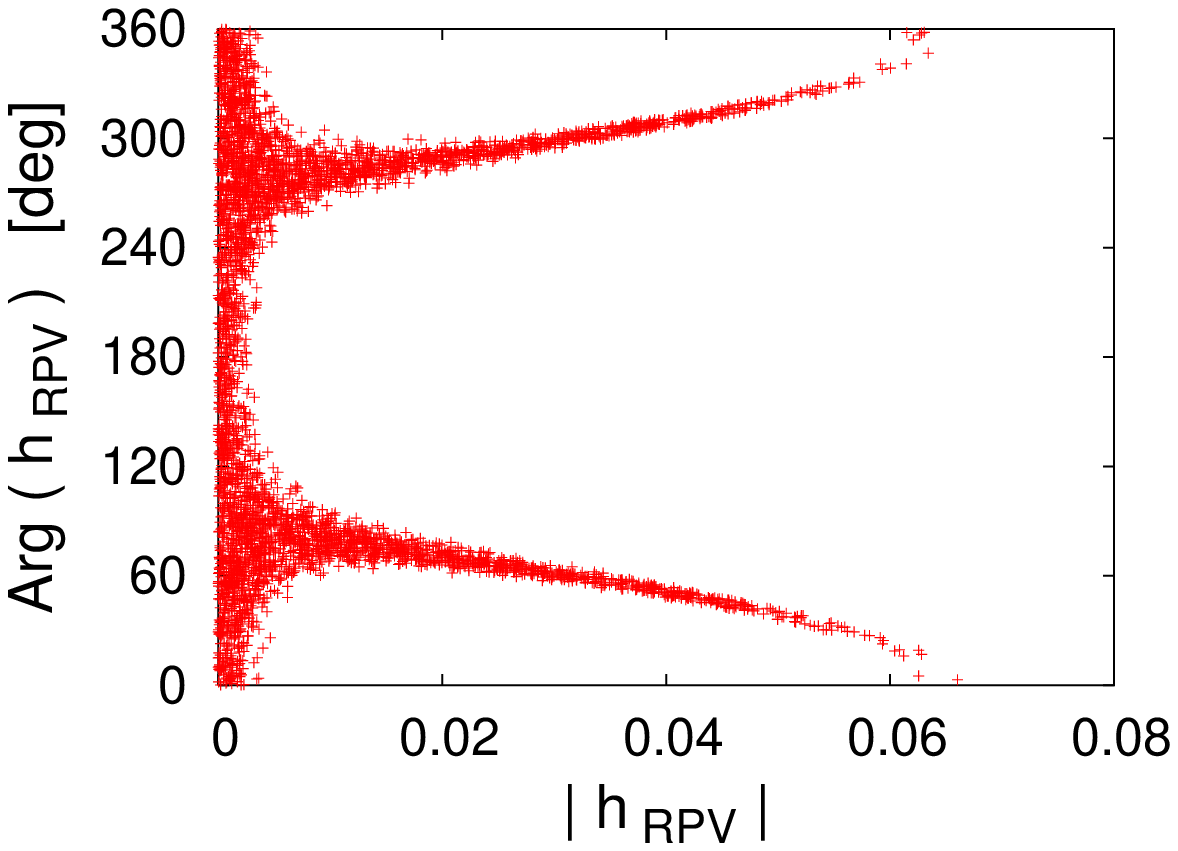,width=8cm,angle=0}
}
\parbox{8cm}{
\epsfig{file=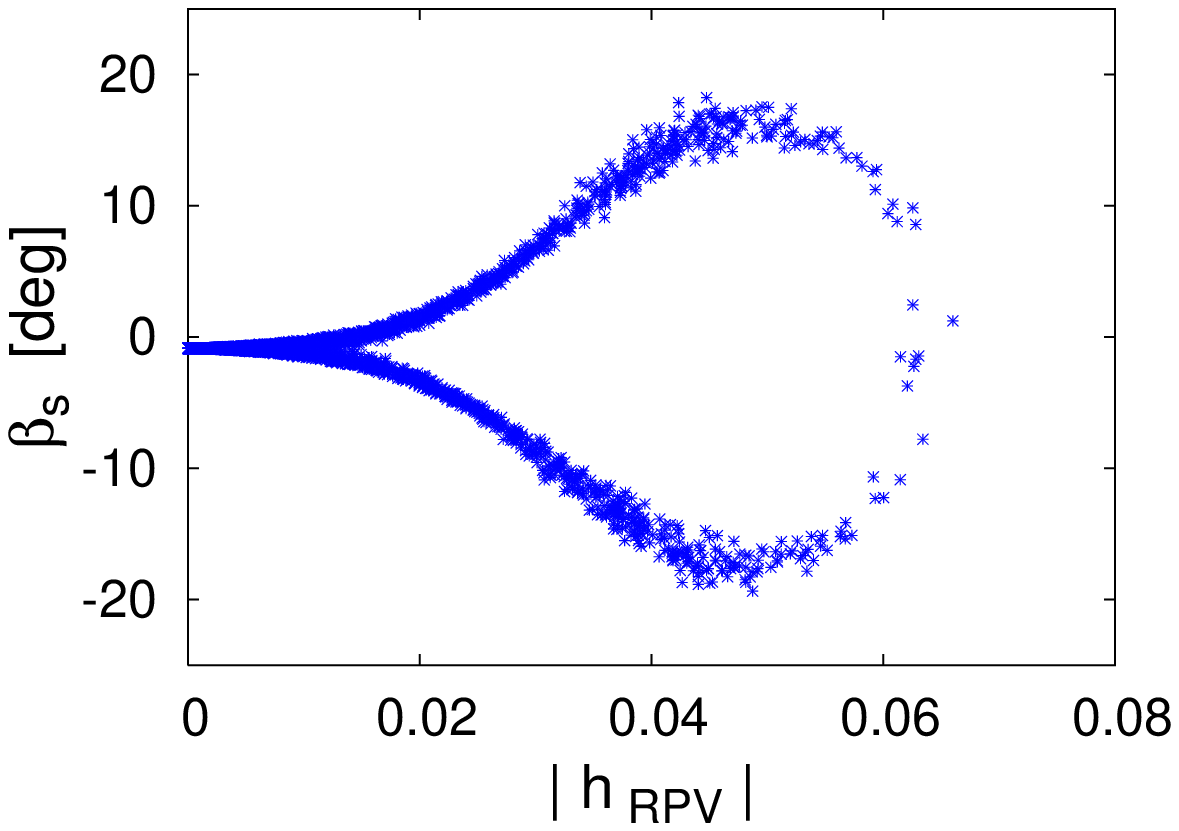,width=8cm,angle=0}
}
\caption{The left panel shows the allowed parameter space of
the phase and the magnitude of the coupling product
$h_{\rm RPV}(b \to s)$.
The right panel shows the allowed values of $\beta_s$ 
as a function of $|h_{\rm RPV}(b \to s)|$.
The masses of the right handed squarks $\tilde t_R$ and $\tilde d_R$
have been taken to be 300 GeV. 
\label{rpv-bs}}
\end{figure}

The R-parity violating couplings that may enhance $\dgq/\gq$ 
also tend to enhance the branching ratios
$B_d \to \phi \pi^0$ and $B^+ \to K^0 \pi^+, \phi \pi^+$.
The relevant measurements are \cite{hfag}
\bea
{\rm BR}(B^+ \to \phi\pi^+) < 0.24\times 10^{-6} \; ,
& \quad \quad &
{\rm BR}(B^+ \to  K^0\pi^+) = (23.1\pm 1.0) \times 10^{-6} \; ,
\nonumber\\
{\rm BR}(B^0 \to \phi \pi^0 ) < 0.28\times 10^{-6} \; .
\label{br-bkpi}
\eea
Taking all squarks to be degenerate at 300 GeV, the 
values of $\dgs/\gs$ and $\beta_s$ consistent with the above
branching ratios and the $\Delta M_s$ measurement 
are shown in Fig.~\ref{rpv-bs}. 
Note that for the form factor in the $B^+ \to K^0 \pi^+$
mode, we use the value of the form factor that reproduces
the central value of the ${\rm BR}(B^+ \to  K^0\pi^+)$
measurement in (\ref{br-bkpi}) with naive factorization,
i.e. we assume that the current measurement shows no
new physics effects.
This makes our new physics estimates rather conservative.

Clearly, the value of $\dgs/\gs$ in this scenario does not
increase much beyond its SM value. However, the CP violating
phase $\beta_s$ may be enhanced to as much as $\pm 20^\circ$,
making the CP violation observable through the decay modes
$B_s \to J/\psi \phi$, 
$B_s \to D_s^{+(*)} D_s^{-(*)}$, 
or $B_s \to J/\psi \eta^{(')}$ \cite{durham}. 

Note that at large $|h_{\rm RPV}|$, values of 
${\rm Arg}(h_{\rm RPV}) \approx 0$ are preferred, whereas
${\rm Arg}(h_{\rm RPV}) \approx \pi$ is strongly 
disfavoured. This indicates that for 
${\rm Arg}(h_{\rm RPV}) \approx 0$ ($\pi$), the SM and
RPV amplitudes to $\Delta M_s$ interfere destructively
(constructively). 
This is because the effective four-Fermi
Hamiltonian for the R-parity violation model comes with an
opposite sign to that of the SM Hamiltonian. 
This may be explained as follows.
The $(S-P)\times (S+P)$ interaction in the RPV SUSY gives 
$-(1/2) (V+A)\times (V-A)$ under Fierz reordering, 
and when the charge-conjugated spinors are replaced by
ordinary spinors, a flip of their positions, which also
changes the $V-A$ Lorentz structure to $V+A$, gives
another negative sign.
On top of that, the internal propagator is scalar and not a 
vector gauge particle, which introduces another negative sign
in the amplitude. The BSM contribution thus has a net
negative sign compared to the SM value, when the relative
weak phase between the two amplitudes is zero.

The RPV SUSY scenario does not give rise to any significant
new physics effects in the $\bdbdbar$ system, since the CP 
violating phase $\beta_d$ is also well measured, and 
in combination with the limit on BR$(B \to \phi \pi)$, 
constrains the new physics parameter space severely.

\section{Summary and conclusions}
\label{summary}

We have studied models with (i) scalar leptoquark and (ii) R-parity
violating supersymmetry as examples of BSM physics that may
enhance the values of the lifetime differences $\dgs/\gs$ as well
as $\dgd/\gd$. Such an enhancement is possible since 
the new couplings in these models are flavour dependent,
and there are additional light degrees of freedom in the
$\bqbqbar$ mixing box diagram. 
The latter contribute to $\Gamma_{12q}$, thus enabling 
the evasion of the so-called Grossman bound,
so that the lifetime differences can be higher.
Since the values of $\dgq/\gq$ in the SM have uncertainties of 
$\sim 30\%$, we look for enhancements of ${\cal O}(1)$,
so that such a large $\dgq/\gq$ would be a clean signal of
new physics.

We summarise our results in Table~\ref{tab-limits}.
Let us point out the salient features that emerge from 
our analysis.


\begin{table}[htbp]
\begin{center}
\begin{tabular}{|l|c|c|c|l|}
\hline
Model& $|h_{\rm BSM}|$ bound & $\Delta\Gamma_q/\Gamma_q$ range & 
$\beta_q$ range & \ Related branching ratios \\
\hline 
$\bsbsbar$ & & & & \\
LQ, 2nd gen. & $ 6 \times 10^{-4} $ & $(-0.01,0.35)$ & $\approx -1^\circ$
 & \ BR$(B_s\to\mu^+\mu^-)< 7.5 \times 10^{-8}$ \\
LQ, 3rd gen. & $0.07$ & $(-0.01,0.51)$ & $ (-45^\circ,20^\circ)$
 & \ BR$(B_s\to\tau^+\tau^-) \to 18\%$ \\
RPV SUSY & $0.065$   &  $(-0.01,0.35)$ & $(-20^\circ,20^\circ)$ & 
\ BR$(B \to K \pi) \approx 2.3 \times 10^{-5}$ \\
\hline
$\bdbdbar$ & & [in \%] & & \\
LQ, 2nd gen. & $4\times 10^{-4}$ & $(0.1,0.5)$  & 
$\approx \beta_d$ (exp) & 
\ BR$(B_d \to \mu^+ \mu^-) < 2.3 \times 10^{-8}$ \\
LQ, 3rd gen. & $0.012$ & (-0.5,2.5) &$\approx \beta_d$ (exp) &
\ BR$(B_d \to\tau^+\tau^-) \to 0.4 \%$ \\
RPV SUSY & $3.2\times 10^{-3}$   & $(0.1,0.5)$ & 
$\approx \beta_d$ (exp) & 
\ BR$(B_d \to \phi \pi) < 0.25 \times 10^{-6}$ \\
\hline
\end{tabular}
\label{paramspace}
\label{tab-limits}
\caption{Allowed values of the magnitude of the coupling product, 
$|h_{\rm BSM}|$, in the models considered in the text.
The weak phase of  $h_{\rm BSM}$ is taken to vary in the entire range 
0--$2\pi$.
We also give the allowed values of the lifetime differences and 
CP phase alongwith the decay modes whose rates have strong
correlations with the enhancement of the lifetime difference.
}
\end{center}
\end{table}

\begin{itemize}
\item
Among the models considered here, 
only the third generation leptoquark model predicts a large enhancement of
$\Delta\Gamma_q/\Gamma_q$ over the SM range. 
In this model, the value of $\dgs/\gs$ can be as high as 0.51,
which is restricted by the current experimental 95\% C.L. bound
from direct measurements.
Improvements in the $\dgs/\gs$ measurements will hence either
detect new physics of this kind, or will bound the coupling product
$\lambda^*_{32} \lambda_{33}$ in this model.

\item 
An enhancement of a factor of up to 5 
is also possible for $\dgd/\gd$ in 
the models with the third generation leptoquark. Whereas the SM
prediction for this quantity is $\sim 0.4\%$, making its
measurement extremely difficult and rather unappealing, 
$\dgd/\gd \sim 2.5\%$ is within the capability of the $B$ factories.
Limiting the value of $\dgd/\gd$ from above also translates to
bounding the coupling product $\lambda^*_{31} \lambda_{33}$ 
in this model.

\item
The enhancement in $\dgq/\gq$ is correlated with an enhancement of
$B_q \to \tau^+ \tau^-$. 
Thus, looking for $\tau$ pairs in BaBar, Belle, and LHCb is of 
major importance. 
With the current constraints on all the parameters in the 
third generation leptoquark model, the branching ratio of 
$B_s \to \tau^+ \tau^-$ ($B_d \to \tau^+ \tau^-$) can be
as high as 18\% (0.4\%).  
Though an extremely difficult measurement, the gains from the
analysis of this decay are significant.

\item The mixing-induced phase $\beta_s$ 
of the $\bsbsbar$ system can be large
with the third generation leptoquark as well as with the RPV SUSY:
$|\beta_s|$ can become as high as $45^\circ$ in the leptoquark
model and up to $20^\circ$ in RPV SUSY. 
Currently there is almost no constraint on $\beta_s$, but improved
measurements, if they do not detect new physics through 
CP-violating signals in $B_s\to J/\psi\phi$, 
$B_s \to D_s^{+(*)} D_s^{-(*)}$, or $B_s \to J/\psi \eta^{(')}$,
can further squeeze the parameter space for BSM physics.

\item Clearly, while non-MFV models are essential for enhancing
$\dgq/\gq$, not all of them serve the purpose.
Among the models we have considered, while the third generation
leptoquark models give large $\dgq/\gq$ enhancement, 
the second generation 
leptoquark models fail to do so because of severe constraints
on their parameter space. The RPV SUSY model, on the other hand,
does not have as tightly constrained couplings, but the 
structure of the new physics contributions is such that though
$\beta_s$ can be enhanced, the value of $\dgq/\gq$ cannot be 
increased. The three scenarios considered by us thus represent
three different facets of non-MFV models.

\end{itemize}

Our analysis uses naive factorisation, and computes $\Gamma_{12q}$
only to leading order. 
In RPV SUSY, we neglect the renormalisation group running of the 
Wilson coefficients and the mixing of the RPV operator with the others.
However, since we claim $\dgq/\gq$ enhancements of up to a 
factor of 5, and since there are anyway uncertainties in the
model predictions due to the unknown masses of 
leptoquarks or squarks, improved
calculations that take care of the above lacunae are not expected
to change the conclusions qualitatively.
We also take the strong phase difference between the SM
and the BSM contributions to the hadronic decays to be vanishing.
However, since the CP violation in hadronic decays does not
play any role in our analysis, and we fit the hadronic decay
form factors to the measured rates, this assumption has no
impact on the predicted enhancement of $\dgq/\gq$.

 Large values of $\dgq$ are possible only in a special class of models
that have flavour dependent couplings and light degrees of freedom
in the $\bqbqbar$ mixing box diagram. In addition, the enhancement
in $\dgq/\gq$ comes coupled with an enhancement in some
correlated decay rates and CP asymmetries. Hence, a measurement of
large $\dgq$ would point one towards very specific sources of
new physics. 
Efforts towards the measurement of $\dgs$ and $\dgd$, as well
as the decay rates like $B_{s/d} \to \tau^+ \tau^-$ are
therefore highly encouraged; they may open the door to
a rich phenomenology.

\section*{Acknowledgements}

AD would like to thank S. Dugad for useful discussions.
AK acknowledges support from the project SR/S2/HEP-15/2003 
of the Department of Science and Technology, Govt.\ of India. 
He also acknowledges discussion with
Emmanuel A. Paschos during the initial stages of the project.

\end{document}